\documentclass[aps,prb,twocolumn,groupedaddress,showpacs,floatfix,altaffilletter]{revtex4-1}

\usepackage{graphicx}
\usepackage{grffile}
\usepackage{subfigure}
\usepackage{amsmath}
\usepackage{amssymb}
\usepackage{bm}
\usepackage{nicefrac}
\usepackage{color}
\usepackage[dvipsnames]{xcolor}
\usepackage{amsmath,amssymb,amsfonts}
\usepackage{epsfig}
\usepackage{times}
\usepackage[colorlinks,bookmarks=false,citecolor=blue,linkcolor=red,urlcolor=blue]{hyperref}

\newcommand{\fref}[1]{Fig.~\ref{#1}}
\newcommand{\eref}[1]{Eq.~(\ref{#1})}

\newcommand{\normwidth}{0.8\columnwidth}
\newcommand{\smallwidth}{0.6\columnwidth}

\begin{document}

\title{Electric polarization in correlated insulators}

\author{R.~Nourafkan, and G.~Kotliar}
\date{\today}

\affiliation{Department of Physics \& Astronomy, Rutgers University, Piscataway, NJ 08854-8019, USA}
\begin{abstract}

We derive a formula for the electric polarization of interacting insulators, expressed in terms of
the full Green's and   vertex functions. We  exemplify this method in the half-filled ionic Hubbard model treated within   dynamical mean field theory (DMFT).  The  electric polarization of a correlated band insulator is determined by the interplay of ionicity and covalency, and both quantities are renormalized by 
the electron-electron interactions. We introduce  quasiparticle approximation to the exact equation for the polarization, and compare the results of this approximation with those of the exact  DMFT formulation and of static  mean field theories such as the LDA+ U. The latter  overestimates the electronic 
contribution to the electric polarization when the quasiparticle weight of the active bands is
very small. 
 \end{abstract}


\maketitle
\section{Introduction}
The electric polarization, ${\bf P}$,  is a measure of   differences in position  between the
center of mass of the  band electrons and the lattice ions. It is generally non zero for a
material which lacks inversion symmetry and plays the role of the order parameter
in the theory of ferroelectricity. Over the past two decades, there 
have been important conceptual and computational advances in the first principles calculation of this quantity. \cite{Resta07}

The modern theory of polarization expresses  ${\bf P}$  in terms of the Berry phase
acquired by a  Slater determinant, describing the  insulating state as an effective
one particle theory. \cite{King-Smith93}  
A complementary picture of the polarization, as the displacement of the
Wannier centers of the single particle states involve, emerges naturally in this formalism. \cite{Vanderbilt93} 

The Berry phase  formalism was extended beyond an effective single particle picture, 
by considering the changes of the  phase  of the many body function of all the electrons in the solid
as a response to the changes in the boundary conditions. \cite{Ortiz94, Resta98} The resulting expression can also be recast in terms of the resolvent of the Hamiltonian of all
the interacting electrons in the many body ground state. \cite{Ortiz}
The many body wave function formalism to calculate polarization was
applied to simple model Hamiltonians, \cite{Wilkens01} but is prohibitively difficult to carry out in practice for realistic models of the electronic structure of a solid. 

In this paper we  formulate the problem of the calculation of the electronic polarization 
of a correlated electronic system in terms of the  one particle  Green's function and vertex functions.
We derive  a general   expression for the polarization, and discuss its implementation 
within  dynamical mean field theory (DMFT). \cite{Georges96} We apply the formalism to a simple model
of a correlated insulator.  The approach  has similarities to recent studies of incorporating correlations in the calculation of the topological indices of topological insulators \cite{Wang10, Go12}, 
and to the early work of Volovik \cite{Volovik03} and gives useful insights into  the effects of   Hubbard correlations
on the electric polarization.

\section{Derivation}

In an extended system 
the change in polarization $\Delta {\bf P}$ induced by distorting an atomic coordinate ${\bf R}_{i\alpha}\rightarrow {\bf R}_{i\alpha}+\xi(t)\Delta{\bf R}_{i\alpha}$ that breaks inversion symmetry is given by the integrated bulk transient current as the system
adiabatically evolves from the initial state ($t=0$) to the final state ($t=T$), i.e.,
\cite{Resta07}

\begin{eqnarray}
\Delta {\bf P}=\int_0^T dt \frac{1}{V_{\rm cell}}\int_{\rm cell}d{\bf r} {\bf J}({\bf r},t)
=\int_0^T dt{\bf J}({\bf q}={\bm 0},t)
\label{eq:pol1}
\end{eqnarray}
where ${\bf J}({\bf r},t)$ denotes the current density and $V_{\rm cell}$ is the primitive-cell volume. Implicit in the analysis is that the system must remain insulating everywhere along the path, as otherwise the adiabatic condition fails. 

The induced current in a small time interval $[t,t+\delta t]$ can be obtained as follows; assume the system is described by $\hat{H}(\xi(t))$ at time $t$. At a later time $t+\delta t$, the distortion of the system changes, i.e.,  $\xi\rightarrow \xi+\delta\xi$, which perturbs the Hamiltonian of the system at time $t$ by $(\delta \hat{H}/\delta\xi)\delta\xi\equiv \hat{H}_{\xi}^{\prime}\delta\xi$ and induces a current. In order to evaluate the current explicitly, we assume $\delta\xi(t)=\delta\xi\exp(-i\nu t+\eta t)$ with $\eta$ an infinitesimal positive number and finally we consider the $\nu \rightarrow 0$ limit. 
Consequently, the response of the system must follow the same time behavior.

\begin{figure*}
\begin{center}
\includegraphics[width=1.5\columnwidth]{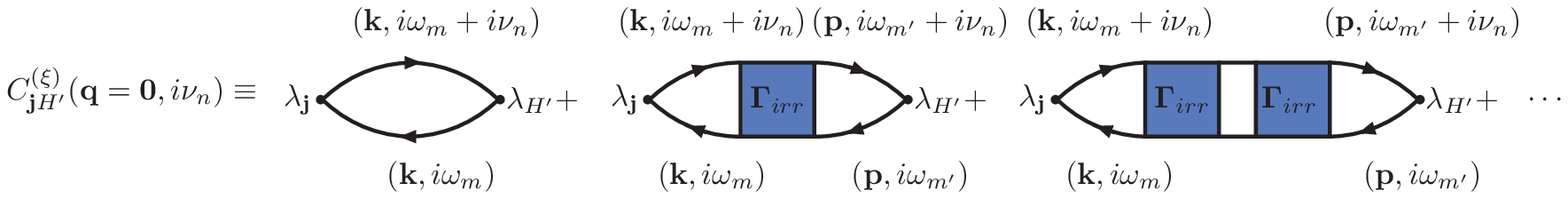}\\
\includegraphics[width=1.7\columnwidth]{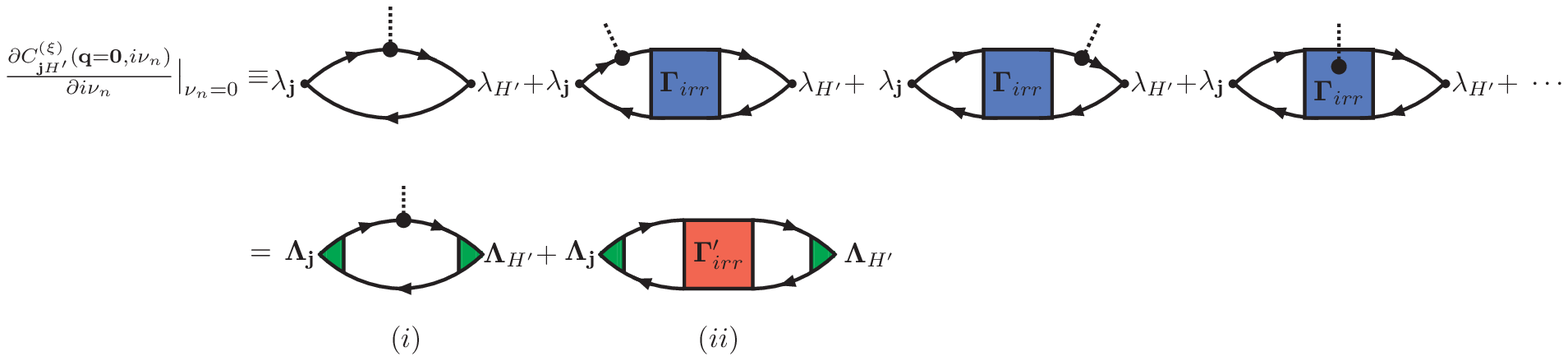}
\caption{(Color online) Diagrammatic expansion of the correlation function $C_{{\bf j}H^{\prime}}^{(\xi)}({\bf q}={\bm 0},i\nu_n)$ and its derivative $(\partial C_{{\bf j}H^{\prime}}^{(\xi)}({\bf q}={\bm 0},i\nu_n)/\partial i\nu_n)|_{\nu_n=0}$. Lines show the full Green's function, ${\bm \Gamma}_{irr}$ is the irreducible particle-hole interaction vertex and ${\bm \Gamma}_{irr}^{\prime} = (\partial{\bm \Gamma}_{irr}/\partial i\nu_n)\big|_{\nu_n=0}$, where $\nu_n$ is bosonic frequency. Dashed line shows $\partial/\partial i\omega_n$ vertex. 
$\bm{\Lambda}_{\bf j}$ and  $\bm{\Lambda}_{H^{\prime}}$ are the particle-hole vertex functions. Bare vertices are given by $\bm{\lambda}^{(\xi)}_{\bf j}({\bf k})=(\partial {\bf H}_0^{(\xi)}/\partial{\bf k})$ and $\bm{\lambda}^{(\xi)}_{H^{\prime}}({\bf k})=(\partial {\bf H}_0^{(\xi)}/\partial \xi)$, respectively.}\label{feyndiag}
\end{center}
\end{figure*}

According to the Kubo formula the induced current density is given by 
\begin{eqnarray}
{\bf J}({\bf q}={\bm 0},t) &=& {\bf J}({\bf q}={\bm 0},\nu)e^{(-i\nu t+\eta t)}\nonumber\\
&=& C^{(\xi)}_{{\bf j}H^{\prime}}({\bf q}={\bm 0},\nu)\delta\xi(t),\label{eq:current01}
\end{eqnarray}
where $C^{(\xi)}_{{\bf j}H^{\prime}}({\bf q}={\bm 0},\nu)$ is ${\bf q}={\bm 0}$ component of the Fourier transform (in space and time) of the retarded correlation function,  \cite{Bruus}
\begin{equation}
C^{(\xi)}_{{\bf j}({\bf r})H^{\prime}({\bf r}^{\prime})}(t-t^{\prime})=-i\Theta(t-t^{\prime})\langle \big[\hat{{\bf j}}_{\xi}({\bf r},t),\hat{H}_{\xi}^{\prime}({\bf r}^{\prime},t^{\prime}) \big]\rangle_{\xi}.
\end{equation}
Here the subscript $\xi$ for operators shows that they are in the Heisenberg representation with respect to the instantaneous Hamiltonian, $\hat{H}(\xi(t))$,  and $\langle \cdots \rangle_{\xi}$ means that the average is taken with respect to the instantaneous spectrum.
In the limit of $\nu=0$ (static distortion), $C^{(\xi)}_{{\bf j}H^{\prime}}({\bf q}={\bm 0},\nu=0)=0$, therefore the driving force is $\delta\dot{\xi}$ and ${\bf J}\propto \delta\dot{\xi}$. In this limit \eref{eq:current01} reduces to  
\begin{multline}
{\bf J}({\bf q}={\bm 0},t) 
=\nu(\frac{\partial}{\partial \nu} C^{(\xi)}_{{\bf j}H^{\prime}}({\bf q}={\bm 0},\nu))\big|_{\nu=0}\delta\xi e^{(-i\nu t+\eta t)}\\
= i(\frac{\partial}{\partial i\nu_n} C^{(\xi)}_{{\bf j}H^{\prime}}({\bf q}={\bm 0},i\nu_n))\big|_{\nu_n=0}\delta\dot{\xi}\equiv{\mathcal L}^{(\xi)}_{{\bf j}H^{\prime}}\delta\dot{\xi}
\label{eq:current1} 
\end{multline}
where we used analytic continuation in the last line and we defined transport coefficient ${\mathcal L}^{(\xi)}_{{\bf j}H^{\prime}}$. Substituting \eref{eq:current1} in \eref{eq:pol1} gives an expression for 
$\Delta{\bf P}=\int_0^1d\xi\;{\mathcal L}^{(\xi)}_{{\bf j}H^{\prime}}$.
Note that $\Re {\mathcal L}^{(\xi)}_{{\bf j}H^{\prime}}$ is antisymmetric under exchange of ${\bf j}$ and $H^{\prime}$ operators, 
i.e.,  $\Re {\mathcal L}^{(\xi)}_{{\bf j}H^{\prime}}=-\Re {\mathcal L}^{(\xi)}_{H^{\prime}{\bf j}}$,  which can be seen from its Lehmann representation (see appendix A).  

\begin{figure}
\begin{center}
\includegraphics[width=0.84\columnwidth]{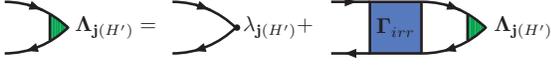}
\caption{(Color online) Bethe-Salpeter equation for the vertex functions, $\bm{\Lambda}_{\bf j}$ and  $\bm{\Lambda}_{H^{\prime}}$, in the particle-hole channel.}
\label{feyndiag2}
\end{center}
\end{figure}

In a diagrammatic series expansion we can express $C^{(\xi)}_{{\bf j}H^{\prime}}$
in terms of the interacting Green's functions and the irreducible particle-hole vertex functions, $\bm{\Gamma}_{irr}$, shown in \fref{feyndiag}.  
We can find the perturbation expansion for ${\mathcal L}_{{\bf j}H^{\prime}}$ by differentiating with respect to $i\nu_n$ of the diagrammatic expansion of $C^{(\xi)}_{{\bf j}H^{\prime}}$. This introduces a derivative vertex $(\partial /\partial i\omega_m)$. 
The  diagrams  contributing at the ${\mathcal L}_{{\bf j}H^{\prime}}$ are of two types: (i)  diagrams which are separated into three pieces, each of which has a vertex, by cutting three electron lines. We denote the contribution of these diagrams to the polarization by $\Delta {\bf P}_1$. (ii) All the rest of the diagrams have  their contribution to the polarization denoted by $\Delta {\bf P}_2$. The change in polarization from the diagrams of type (i) is given by

\begin{multline}
\Delta {\bf P}_1=-\int_0^1 d\xi\;(\frac{ie}{2N\beta})\sum_{{\bf k}\sigma} 
\sum_{\omega_m}
 \\{\rm Tr}\bigg(
\bm{\Lambda}^{(\xi)}_{\bf j}
{\bf G}^{(\xi)}
\frac{\partial{\bf G}^{(\xi)-1}}{\partial {i\omega_m}} 
{\bf G}^{(\xi)}\bm{\Lambda}^{(\xi)}_{H^{\prime}}
{\bf G}^{(\xi)}-\\
\bm{\Lambda}^{(\xi)}_{H^{\prime}}
{\bf G}^{(\xi)}
\frac{\partial{\bf G}^{(\xi)-1}}{\partial {i\omega_m}} 
{\bf G}^{(\xi)}\bm{\Lambda}^{(\xi)}_{\bf j}
{\bf G}^{(\xi)}
\bigg)
\label{eq:pol4}
\end{multline}
where we have used the identity
 $(\partial {\bf G}^{(\xi)}/\partial i\omega_m)=-{\bf G}^{(\xi)}(\partial {\bf G}^{(\xi)-1}/\partial i\omega_m){\bf G}^{(\xi)}$ and the antisymmetric property of ${\mathcal L}_{{\bf j}H^{\prime}}$ in writing \eref{eq:pol4} down. $\bm{\Lambda}_{\bf j}$,  $\bm{\Lambda}_{H^{\prime}}$ are the particle-hole vertex functions that can be obtained from the so-called Bathe-Salpeter equation which is graphically shown in \fref{feyndiag2}. $\beta$ is the inverse temperature.

The interacting single particle Green's function of the system is
\begin{equation}
{\bf G}^{(\xi)}_{{\bf k} \sigma}(i\omega_m)=[(i\omega_n+\mu){\bm 1}-{\bf H}_0^{(\xi)}({\bf k})-{\bm \Sigma}^{(\xi)}_{\sigma}({\bf k},i\omega_m)]^{-1}, \label{eq:gf0}
\end{equation}
where ${\bf H}_0^{(\xi)}({\bf k})$ denotes the noninteracting part of the Hamiltonian. While ${\bf G}^{(\xi)}$ and ${\bf H}_0^{(\xi)}$ are gauge covariant, $\Delta{\bf P}$ is gauge invariant. However, the form of the interaction matrix and the expression of the current do depend on the phase of the tight-binding basis. Here we follow the choice of Ref. \onlinecite{Paul03} where makes the Peierls expression for current more accurate: ${\bf H}_0$ satisfies ${\bf H}_0({\bf k+\bf K})=O^{\dagger}{\bf H}_0({\bf k})O$, where $O$ is a diagonal matrix with $O_{nn}=\exp(i{\bf K}.{\bf d}_n)$,  $n$-th orbital is located at ${\bf d}_n$, and $\bf K$ is the reciprocal lattice vector. \cite{periodicgauge} ${\bm \Sigma}^{(\xi)}_{ \sigma}({\bf k},i\omega_m)$ denotes the electron self-energy. Bold quantities are $n\times n$ matrices where $n$ denotes number of orbitals within the unit cell.

The diagrams of type (ii) do not contribute to an antisymmetric transport coefficient. \cite{Ishikawa87}  Therefore $\Delta {\bf P}_2 = 0$ [see appendix B for a discussion on the type (ii) diagrams]. 

One can use the Ward identity to replace the electric current vertex by $\bm{\Lambda}^{(\xi)}_{\bf j}({\bf k})=-(\partial {\bf G}^{(\xi)}_{{\bf k} \sigma}(i\omega_m)^{-1}/\partial {\bf k})$ and the distortion current vertex by $\bm{\Lambda}^{(\xi)}_{H^{\prime}}({\bf k})=-(\partial {\bf G}^{(\xi)}_{{\bf k} \sigma}(i\omega_m)^{-1}/\partial {\xi})$. \cite{Ishikawa87} Therefore, the change in the electric polarization
 can be rewritten as

\begin{eqnarray}
\Delta &{\bf P}&=-\int_0^1 d\xi\;(\frac{ie}{N\beta})\sum_{{\bf k}\sigma} 
\sum_{\omega_m}
\frac{1}{6}\varepsilon^{{\bf k}\xi \omega_m}
\nonumber \\&{\rm Tr}&\bigg(
\frac{\partial{\bf G}^{(\xi)-1}}{\partial {\bf k}}
{\bf G}^{(\xi)}
\frac{\partial{\bf G}^{(\xi)-1}}{\partial {i\omega_m}} 
{\bf G}^{(\xi)}\frac{\partial{\bf G}^{(\xi)-1}}{\partial \xi}
{\bf G}^{(\xi)}
\bigg)
\label{eq:pol5}
\end{eqnarray}
where we have introduced an antisymmetric tensor, $\varepsilon^{{\bf k}\xi \omega_m}$. Equation \ref{eq:pol5} follows from \eref{eq:pol4} due to the antisymmetric property of the derivative relative to ${\bf k}$ and $\xi$ and cyclic property of the trace. In the noninteracting case \eref{eq:pol5} reduces to the Berry phase formula (see appendix C). \cite{King-Smith93}

We emphasis here that \eref{eq:pol5} is an exact formula and can be used along with any method capable of  calculating the interacting Green's function, like GW or DMFT. In the following we will work in the DMFT approximation, where we keep only the self-energy corrections associated with electron correlation (In the appendix D we show that the current vertex corrections are identically zero for the model Hamiltonian considered here). In DMFT, the correlation function reduces to the bubble contribution with the fully interacting Green's function and the following vertices: 
$(\partial{\bf G}^{(\xi)-1}/\partial {\bf k}) = -(\partial {\bf H}_0^{(\xi)}/\partial{\bf k})$, 
$(\partial{\bf G}^{(\xi)-1}/\partial {\xi})= -(\partial {\bf H}_0^{(\xi)}/\partial{\xi})-(\partial {\bm \Sigma}^{(\xi)}/\partial{\xi})$, and  $(\partial{\bf G}^{(\xi)-1}/\partial {i\omega_n})= {\bm 1}-(\partial {\bm \Sigma}^{(\xi)}/\partial{i\omega_n})$.

\section{ionic Hubbard model} 
 
To investigate the influence of Hubbard correlations on the electronic part of the polarization 
and to benchmark our formalism we turn to a simple model of ferroelectric materials, the ionic Hubbard model (IHM) in one dimension. On a bi-partite lattice the IHM Hamiltonian is defined by the following equation \cite{Giovannetti09, Egami93, Resta95, Freo02}:
\begin{eqnarray}
H=&-&t\sum_{i\sigma}[1+(-1)^i \xi] (c^{\dagger}_{i\sigma}c_{i+1\sigma}+H.c.)\nonumber \\
&+&\Delta\sum_{i \sigma}(-1)^i n_{i\sigma}+U\sum_i (n_{i\uparrow}-\frac{1}{2})(n_{i\downarrow}-\frac{1}{2}),
\label{eq:Ham}
\end{eqnarray}
where, $t(>0)$ and $U$ denote the hopping amplitude between nearest-neighbor sites and the on-site Coulomb repulsion, respectively. 
$-\Delta$($+\Delta$) denotes the local potential energy for the $A$($B$) sublattice, respectively.

The model is relevant for the study of organic ferroelectrics  with stacks of alternating donor and acceptor molecules. 
The paraelectric phase is centrosymmetric (so that all $A-B$ distances are the same), 
whereas in the ferroelectric phase there
is a relative displacement of the $A$ and $B$ sublattices, which
breaks the inversion symmetry and produces an alternating pattern of short and long $A-B$ distances.
Here we modify the hopping amplitude by $\pm\xi$ alternatively
to include a dimer term that breaks the inversion symmetry. 

At half filling, the noninteracting undistorted system ($\xi=0$) is a band insulator (BI) with a charge gap equal to $\delta_c=2\Delta$. Upon turning on and increasing $U$ the charge gap, $\delta_c$, in the correlated band insulator shrinks until a discontinuous transition to a Mott insulator (MI) occurs with a hysteresis region, if the antiferromagnetic (AF) long-range order is not allowed to set in. In the Mott phase a large gap is established. \cite{Garg06, Craco08} As the system is driven deeper into the Mott phase, the gap size increases. Both the BI and the MI are nonpolar at $\xi=0$; however, depending on how one defines the zero reference point of the polarization one can assign a polarization value to these states. Here we define the zero reference point to be the usual band insulator where both electrons occupy the Wannier function centered on the lower energy site, then using the classical point charge model we assign $P^{(\rm el)}/ea=-1/2$ to a Mott insulator with no long range order. In the centrosymmetric system without spontaneous inversion symmetry breaking \cite{BO}, increasing $U$ from zero causes a charge exchange between sites. However, charge flows symmetrically, the macroscopic current is zero and therefore $\Delta P^{(\rm el)}=0$. In the presence of the AF order, the system has a transition to the AF phase at a smaller critical $U$. \cite{Byczuk09}

\begin{figure}
\begin{center}
\includegraphics[width=\normwidth]{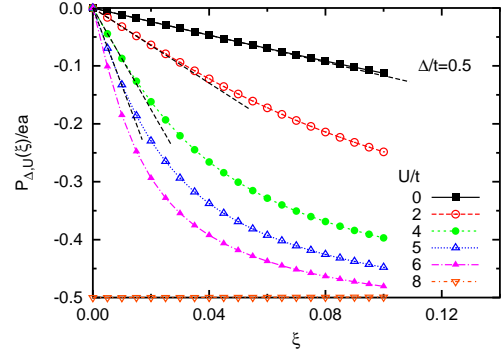}
\caption{(Color online) Electric polarization as a function of distortion $\xi$ for several value of $U$ in the PM phase. For $U/t<7.5 $ the system is in the band insulator phase with zero reference point of polarization $P(\xi=0)/ea=0$. For $U/t=8.0$ data is shown in the Mott phase with zero reference point of polarization $P(\xi=0)/ea=-1/2$. Black dashed lines show \eref{eq:lp}. }\label{fig2}
\end{center}
\end{figure}

Next we present our results in the presence of a distortion. We set $\Delta=0.5t$ and use \eref{eq:pol5} and DMFT to evaluate the electronic part of the electric polarization of the distorted IHM 
in the paramagnetic (PM) and AF phases. 
The Bloch representation of the noninteracting part of the IHM Hamiltonian, \eref{eq:Ham}, is given by 
\begin{equation}
{\bf H}_0(k)=-2t\cos(ka/2){\bm \tau}_x+ 2\xi t\sin(ka/2){\bm \tau}_y-\Delta{\bm \tau}_z
\label{eq:ham0}
\end{equation}
where ${\bm \tau}_{x,y,z}$ are the Pauli matrices and $a$ is unit cell length. In the DMFT approximation the current vertex and the distortion vertex are given by  
\begin{eqnarray}
\frac{\partial {\bf G}^{(\xi)-1}}{\partial k}&=&at\sin(ka/2){\bm \tau}_x+a\xi t\cos(ka/2){\bm \tau}_y,\label{eq:jvertex}\\
\frac{\partial {\bf G}^{(\xi)-1}}{\partial \xi}&=&2 t\sin(ka/2){\bm \tau}_y-
\frac{\partial {\bm \Sigma}^{(\xi)}}{\partial \xi}.\label{eq:Hvertex}
\end{eqnarray}
where ${\bm \Sigma}$ is a diagonal matrix with elements $\Sigma_A$ and $\Sigma_B$. 

Figure \ref{fig2} shows the polarization of the system, $P^{(\rm el)}_{\Delta,U}(\xi)/ea$,  obtained from an evaluation of  \eref{eq:pol5} for several values of $U$ as a function of $\xi$. As can be seen in the correlated BI ($U < U_c\simeq 7.5t$), the absolute value of the $P^{(\rm el)}_{\Delta,U}(\xi)/ea$ increases with $\xi$ and saturates at $-1/2$. The saturation value happens at smaller $\xi$ upon increasing $U$ (note that for small values of $U$ the saturation value is not apparent on this figure, as it occurs at larger $\xi$). 
In the noninteracting case, a system with small ionicity is more polarizable. 
In order to explain results with nonzero $U$ with $U<U_c$, one needs to understand how the correlation renormalizes the ionicity and the covalency of the system. To get some insight we work in the QP approximation and derive an analytical equation for the electric polarization at small dimerization. 
The QP Green's function is obtained by linearizing \eref{eq:gf0} at small $i\omega_m$, 
\begin{eqnarray}
{\bf G}^{(\xi)}_{qp}({\bf k},i\omega_m)&=&{\bf z}^{1/2}[i\omega_m{\bm 1}-{\bf H}^{(\xi)}_{qp}({\bf k})]^{-1}{\bf z}^{1/2},\label{eq:qpgf2}
\end{eqnarray}
where ${\bf H}^{(\xi)}_{qp}({\bf k})={\bf z}^{1/2}({\bf H}^{(\xi)}_0({\bf k})+\Re {\bm \Sigma}^{(\xi)}_{U}(0)-\mu {\bm 1}){\bf z}^{1/2}$ is the QP Hamiltonian. The quasi-particle weight is given by ${\bf z}=[{\bm 1}-\Im {\bm \Sigma}(i\omega_0)/\omega_0]^{-1}$,
where $\omega_0=\pi/\beta$ is first fermionic Matsubara frequency. In our case ${\bf z}$ is a diagonal matrix with $ z_A=z_B\equiv z$. Substituting \eref{eq:qpgf2} into  \eref{eq:pol5}, using the fact that $\partial {\bf G}_{qp}^{-1}/\partial i\omega_n={\bf z}^{-1}$ and following the derivation presented in  
appendix C, one can see that in the QP approximation the polarization of the IHM for small distortions is given by
\begin{equation}
P^{(\rm el)}_{\Delta,U}(\xi) \simeq -\frac{2ea}{N}\sum_k\frac{(\Delta_{qp}^{ren}/(zt))\sin^2(ka/2)}{[(\Delta_{qp}^{ren}/(zt))^2+4\cos^2(ka/2)]^{3/2}}\xi,
\label{eq:lp} 
\end{equation}
which shows that for small $\xi$ the electric polarization of the interacting system is obtained by replacing bare quantities with renormalized ones, $\Delta\rightarrow \Delta_{qp}^{ren}$ and $t\rightarrow zt$.

The renormalized ionicity can be obtained from the renormalized charge gap which in turn can be obtained by identifying the poles of the renormalized propagator, \eref{eq:qpgf2}, and is given by
\begin{eqnarray}
\delta_{c}^{qp}=2\Delta_{qp}^{ren}&=& 2z[\Delta-\Re( \Sigma_{A}(i\omega_0)- \Sigma_{B}(i\omega_0))].
\label{eq:rengap}
\end{eqnarray}

The frequency range of the validity of the Fermi liquid assumptions decreases upon increasing $U$, but the charge gap also shrinks simultaneously. A closer investigation of the self-energy and local density states of the correlated insulator show that one can continue to assume the Fermi liquid assumptions on the frequency range between the highest pole of the valence band and the lowest pole of the conduction band for all interaction strengths. Therefore, \eref{eq:qpgf2} describes low energy physics precisely and \eref{eq:rengap} is the true charge excitation gap, $\delta_{c}=\delta_{c}^{qp}$. 

Upon increasing $U$, $(\Delta_{qp}^{ren}/t)$ decreases faster than $z$ which leads to a reduction of $(\Delta_{qp}^{ren}/(zt))$ which in turn leads to a more polarizable system where a small distortion can trigger substantial changes in the electric polarization. In \fref{fig2} the QP result \eref{eq:lp} is shown with dashed lines. The agreement between the QP approximation and the full answer is very good in the linear regime of polarization.  

\begin{figure}
\begin{center}
\includegraphics[width=\normwidth]{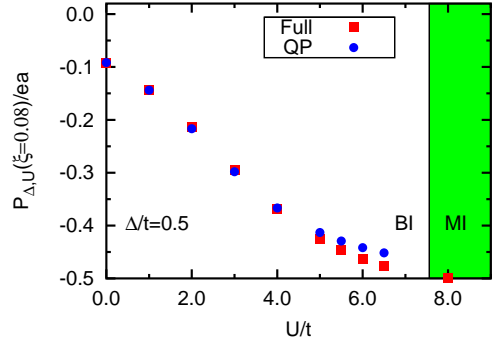}
\caption{(Color online) Electric polarization for a distorted system with $\xi=0.08t$ for several interaction strengths, $U$, using the full calculation and the quasi-particle (QP) approximation.  }\label{fig2.5}
\end{center}
\end{figure}

Figure \ref{fig2.5} shows $P^{(\rm el)}_{\Delta,U}(\xi=0.08t)/ea$ for several values of $U$ obtained from the full and QP calculations. In the QP calculation we employ ${\bf H}^{(\xi)}_{qp}$ to define an effective noninteracting system and calculate $P$ using the Berry phase formula. \cite{King-Smith93} It is worth mentioning that the same results for polarization can be obtained by working with $({\bf H}^{(\xi)}_0({\bf k})+\Re {\bm \Sigma}^{(\xi)}_{U}(0)-\mu {\bm 1})$ as the QP Hamiltonian, which provides the correct ratio of the renormalized ionicity to the covalency. However, the latter Hamiltonian gives incorrect results on the charge gap and does not account for quasi-particle weight renormalization. Consistent with \eref{eq:lp}, the QP results agree very well with the full calculations at weak to intermediate interaction strengths. However, at strong coupling, $P$ obtained from the QP method deviates from the full answer because it does not account for the bandwidth renormalization correctly. 

At large interaction strength, for example $U=8.0t$, the system is in a Mott phase. In a Mott phase, the variation in the electronic part of the polarization as a function of the distortion is very small (here we restrict ourself to small $\xi$). Indeed in a Mott phase, charge fluctuations are strongly suppressed and an ionic displacement induces only a very small current. In the Mott phase the effective noninteracting system described by  ${\bf H}_{qp}$ is an extremely ionic system where one sublattice is fully occupied and the other one is empty. In reality, however, in the true ground state the charge is almost uniformly distributed in the system. Thus, the QP approximation fails in the Mott phase.

\begin{figure}
\begin{center}
\includegraphics[width=\smallwidth]{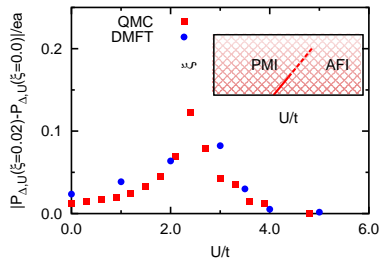}
\caption{(Color online) $|P^{(\rm el)}_{\Delta,U}(\xi=0.02)-P^{(\rm el)}_{\Delta,U}(\xi=0)|$ obtained from an implementation of \eref{eq:pol5} along with the QMC calculation reproduced from \cite{Wilkens01}. For $U<2.4t$ the system is in the PM phase. For larger $U$ it is in AF phase. Inset shows the DMFT schematic phase diagram of the system. }\label{fig3}
\end{center}
\end{figure}

Next we investigate the AF phase which allows us to compare our results with the QMC results on IHM. Investigation of the staggered magnetization, $m=\langle n_{A\uparrow}-n_{A\downarrow}\rangle=-\langle n_{B\uparrow}-n_{B\downarrow}\rangle$, of the centrosymmetric structure, $\xi=0$, as a function of the interaction strength shows that at zero temperature the system shows long-range AF for $U \geq 2.4t$. Figure \ref{fig3} shows $|P^{(\rm el)}_{\Delta,U}(\xi=0.02)-P^{(\rm el)}_{\Delta,U}(\xi=0)|$ obtained from an implementation of \eref{eq:pol5}  along with the QMC calculation reproduced from \cite{Wilkens01}. In the QMC data, the phase diagram of the undistorted ionic Hubbard model includes a spontaneous dimerized phase. In DMFT the onset of the AF phase coincides numerically with the onset of the spontaneous bond order phase in the QMC. \cite{Kancharla07}  
The two calculations agree very well with one another because both the AF  phase and the spontaneous bond order phase have similar AF short-range correlation and the electric polarization depends only on short range correlation.  

\section{conclusions}

In conclusion, we have introduced a
practical many body approach to the
calculation of the electric
polarization of interacting insulators.
We have implemented and tested the
formalism in the context of a model
Hamiltonian of a correlated band insulator and shown that while
the electronic polarization is affected
by Hubbard correlations, it is a less
sensitive quantity well approximated
by a quasi-particle approximation and
depending on the ratio of the gap
and the covalency of the quasi-particle
band structure. These results
justify the success
of the traditional electronic structure
of methods which do not include an
explicit frequency dependence of the
self energy . Notice however, that in
some correlated materials such as HoMnO$_3$ this method has
been reported to overestimate the
electric polarization. \cite{Picozzi07, Chai12, Stroppa10}
This can be understood as resulting
from the effects of the bandwidth
renormalization on the polarization
described here. The formalism
of this paper can be implemented
in realistic LDA+DMFT codes.
This together with a comparison
with accurate experimental ARPES
studies of correlated ferroelectrics
are interesting open problems.

\begin{acknowledgments}
We are grateful to F. Marsiglio, M. E. Pezzoli, A. Soluyanov, G. Ortiz, K. Haule and D. Vanderbilt for useful discussions. This work has been supported by NSF DMR-0906943 and NSF DMR-1308141.
\end{acknowledgments}

\appendix
\section{Antisymmetric property}\label{App:AppendixA}
The antisymmetric property of the transport coefficient appearing in Eq.~(3) can be seen from the Lehmann representation of the correlation function. In the general case we have for the correlation function \cite{Bruus}

\begin{equation}
C^R_{AB}(\nu)=\frac{1}{Z}\sum_{nm}\frac{\langle n |A|m\rangle \langle m |B|n \rangle}{\nu+E_n-E_m+i\eta}
\left(e^{-\beta E_n}-e^{-\beta E_m} \right)
\end{equation} 
where $Z={\rm Tr} [\exp(-\beta H)]$. Therefore, ${\mathcal L}_{AB}\equiv i(\frac{\partial}{\partial \nu}C^R_{AB})\big|_{\nu=0}$ has the following form

\begin{equation}
{\mathcal L}_{AB}=-i\frac{1}{Z}\sum_{nm}\frac{\langle n |A|m\rangle \langle m |B|n \rangle}{(E_n-E_m+i\eta)^2}
\left(e^{-\beta E_n}-e^{-\beta E_m} \right)
\label{eq:Leh1}
\end{equation} 
Now if we commute $A$ and $B$ operators, we get
\begin{equation}
{\mathcal L}_{BA}=-i\frac{1}{Z}\sum_{nm}\frac{\langle n |B|m\rangle \langle m |A|n \rangle}{(E_n-E_m+i\eta)^2}
\left(e^{-\beta E_n}-e^{-\beta E_m} \right)
\label{eq:Leh2}
\end{equation}
Comparing \eref{eq:Leh1} and \eref{eq:Leh2} gives
\begin{eqnarray}
\Re {\mathcal L}_{AB}&=&-\Re {\mathcal L}_{BA}\\
\Im {\mathcal L}_{AB}&=&\Im {\mathcal L}_{BA}
\end{eqnarray} 
since only real part is giving the current, then one conclude that the transport coefficient is antisymmetric.

\section{Type (ii) diagrams}\label{App:AppendixB}
\begin{figure}
\begin{center}
\includegraphics[width=0.5\columnwidth]{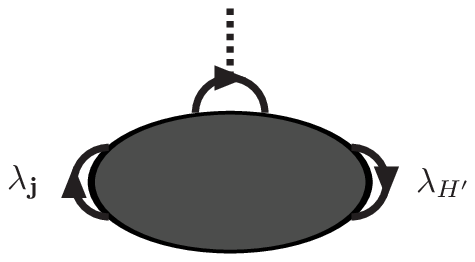}\\
\includegraphics[width=0.5\columnwidth]{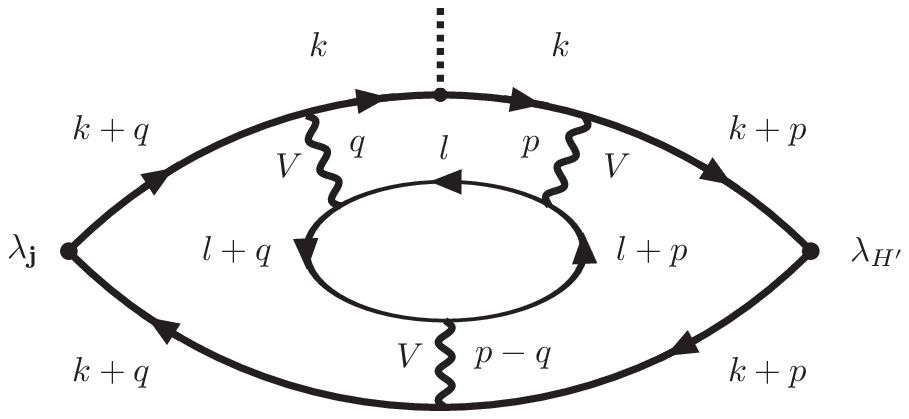}
\caption{First row: General structure of the type (ii) diagrams. 
Second row: The lowest order digram of type (ii) for a system with e-e interaction. Lines show noninteracting Green's function,  wavy lines denote interaction and dashed line shows $\partial/\partial i\omega_m$ vertex. ${ V}(q)$ is bare Coulomb vertex.}
\label{diag1}
\end{center}
\end{figure}

Diagrams that belong to the type (ii) are those that can not be separated into three pieces that have vertices when three fermion lines are cutted. Their contribution in the electric polarization can be written as the third derivative with respect to the vertices (see Fig.~\ref{diag1}, first row). For example, the second row of Fig.~\ref{diag1} 
shows the lowest order diagram of type (ii) for a system with e-e interactions. Lines show the non-interacting Green's functions, wavy lines denote the interaction and the dashed line shows the $\partial/\partial i\omega_m$ vertex.
One can write its 
contribution in the electric polarization as [using the notation $k\equiv({\bf k},i\omega_k)$]
\begin{eqnarray}
&\propto&-i\int_0^1 d\xi\int_0^1 d\xi^{\prime}\delta(\xi-\xi^{\prime} )\sum_{kpql}\sum_{i_1\ldots i_{12}} 
\sum_{\sigma\sigma^{\prime}}
\frac{\partial}{\partial {\bf k}}\frac{\partial}{\partial i\omega_k}\frac{\partial}{\partial \xi}\nonumber\\&\times&
\bigg(g^{(\xi)\sigma}_{i_{12}i_1}(k+q) g^{(\xi)\sigma}_{i_{2}i_5}(k) g^{(\xi)\sigma}_{i_{6}i_{11}}(k+p) 
\nonumber\\&\times&
g^{(\xi^{\prime})\sigma^{\prime}}_{i_{4}i_9}(l+q)V^{i_1i_2\sigma}_{i_3i_4\sigma^{\prime}}(q) 
g^{(\xi^{\prime})\sigma^{\prime}}_{i_{8}i_3}(l)\nonumber\\&\times&
V^{i_5i_6\sigma}_{i_7i_8\sigma^{\prime}}(p)
g^{(\xi^{\prime})\sigma^{\prime}}_{i_{10}i_7}(l+p)V^{i_9i_{10}\sigma^{\prime}}_{i_{11}i_{12}\sigma}(p-q)\bigg)
\label{eq:diag13}
\end{eqnarray}
where ${ V}^{i_1i_2\sigma}_{i_3i_4\sigma^{\prime}}(q)$ is bare Coulomb vertex and for Hubbard model is constant ($=U/2$). Conservation law in the indexes $i_1+i_3=i_2+i_4$ holds. From commutation relation of the derivatives one can see that the type (ii) diagrams are symmetric and they do not have any contribution in the antisymmetric transport coefficient.   

\section{Non-interacting system}\label{App:AppendixC}

Here we show that in the noninteracting case Eq.~(4) reduces to the Berry phase formula. Using the band representation of the Green's function, ${\bf g}^{(b)}_{{\bf k} \sigma}(i\omega_m,\xi)=[i\omega_m{\bm 1}-{\bm \epsilon}^{(\xi)}_{{\bf k}\sigma}]^{-1}$, and $\partial_{i\omega_m} {\bf g}^{(b)}_{{\bf k} \sigma}(i\omega_m,\xi)^{-1}={\bm 1}$ one can rewrite Eq.~(4) as 

\begin{widetext}
\begin{eqnarray}
\frac{d{\bf P}^{(\rm el)}}{d\xi}&=&-\frac{ie}{2N\beta}\sum_{{\bf k}\sigma} 
\sum_{\omega_m}\varepsilon^{{\bf k}\xi}
{\rm Tr}\bigg({\bf g}^{(b)}_{{\bf k} \sigma}(i\omega_m,\xi)
(\frac{\partial {\bf H}^{(\xi)}_{{\bf k} \sigma}
}{\partial {\bf k}})
{\bf g}^{(b)}_{{\bf k} \sigma}(i\omega_m,\xi)
(\frac{\partial {\bf H}^{(\xi)}_{{\bf k} \sigma}
}{\partial\xi}){\bf g}^{(b)}_{{\bf k} \sigma}(i\omega_m,\xi)
\bigg)\nonumber \\ 
&=&-\frac{ie}{2N\beta}\sum_{{\bf k}\sigma} \sum_{n}\sum_{m}
\sum_{\omega_m}\frac{(\frac{\partial {\bf H}^{(\xi)}_{{\bf k} \sigma}
}{\partial {\bf k}})_{nm}(\frac{\partial {\bf H}^{(\xi)}_{{\bf k} \sigma}
}{\partial\xi})_{mn}-(\frac{\partial {\bf H}^{(\xi)}_{{\bf k} \sigma}
}{\partial\xi})_{nm}(\frac{\partial {\bf H}^{(\xi)}_{{\bf k} \sigma}
}{\partial {\bf k}})_{mn}}{(i\omega_m-\epsilon^{(\xi)}_{n{\bf k}\sigma})(i\omega_m-\epsilon^{(\xi)}_{m{\bf k}\sigma})(i\omega_m-\epsilon^{(\xi)}_{n{\bf k}\sigma})}
\nonumber \\ 
&=&\frac{ie}{2N\beta}\sum_{{\bf k}\sigma} \sum_{n}\sum_{m}\frac{\partial}{\partial \epsilon^{(\xi)}_{n{\bf k}\sigma}}
\sum_{\omega_m}\frac{(\frac{\partial {\bf H}^{(\xi)}_{{\bf k} \sigma}
}{\partial {\bf k}})_{nm}(\frac{\partial {\bf H}^{(\xi)}_{{\bf k} \sigma}
}{\partial\xi})_{mn}-(\frac{\partial {\bf H}^{(\xi)}_{{\bf k} \sigma}
}{\partial\xi})_{nm}(\frac{\partial {\bf H}^{(\xi)}_{{\bf k} \sigma}
}{\partial {\bf k}})_{mn}}{(i\omega_m-\epsilon^{(\xi)}_{n{\bf k}\sigma})(i\omega_m-\epsilon^{(\xi)}_{m{\bf k}\sigma})}
\end{eqnarray}
\end{widetext}

By summing over  Matsubara frequency and using the fact that the Fermi function is one for occupied bands and zero for unoccupied bands, we find (terms with $n=m$ vanish)

\begin{widetext}
\begin{eqnarray}
\frac{d{\bf P}^{(\rm el)}}{d\xi}&=&-\frac{ie}{N}\sum_{{\bf k}\sigma} \sum_{n \in occ}\sum_{m\ne n}
\frac{(\frac{\partial {\bf H}^{(\xi)}_{{\bf k} \sigma}
}{\partial {\bf k}})_{nm}(\frac{\partial {\bf H}^{(\xi)}_{{\bf k} \sigma}
}{\partial\xi})_{mn}-(\frac{\partial {\bf H}^{(\xi)}_{{\bf k} \sigma}
}{\partial\xi})_{nm}(\frac{\partial {\bf H}^{(\xi)}_{{\bf k} \sigma}
}{\partial {\bf k}})_{mn}}{(\epsilon^{(\xi)}_{n{\bf k}\sigma}-\epsilon^{(\xi)}_{m{\bf k}\sigma})^2}
\label{eq:pol11}
\end{eqnarray}

which reduces to the Berry curvature \cite{Xiao10}

\begin{eqnarray}
\frac{d{\bf P}^{(\rm el)}}{d\xi}
&=&-\frac{ie}{N}\sum_{{\bf k}\sigma}\sum_{n \in occ} 
\big[\langle \partial_{\xi}u^{(\xi)}_{n{\bf k}\sigma}|
\nabla_{\bf k}u^{(\xi)}_{n{\bf k}\sigma} \rangle -
\langle \nabla_{\bf k}u^{(\xi)}_{n{\bf k}\sigma}|\partial_{\xi}u^{(\xi)}_{n{\bf k}\sigma} \rangle 
\big]\nonumber \\
&=&-\frac{ie}{N}\sum_{{\bf k}\sigma}\sum_{n \in occ} 
\big[\partial_{\xi}\langle u^{(\xi)}_{n{\bf k}\sigma}|
\nabla_{\bf k}u^{(\xi)}_{n{\bf k}\sigma} \rangle -
\nabla_{\bf k}\langle u^{(\xi)}_{n{\bf k}\sigma}|\partial_{\xi}u^{(\xi)}_{n{\bf k}\sigma} \rangle 
\big]
\label{eq:pol12}
\end{eqnarray}
\end{widetext}
using $\langle u^{(\xi)}_{n{\bf k}\sigma}|\frac{\partial {\bf H}^{(\xi)}_{{\bf k} \sigma}
}{\partial{\bf k}}|u^{(\xi)}_{m{\bf k}\sigma} \rangle = (\epsilon^{(\xi)}_{n{\bf k}\sigma}-\epsilon^{(\xi)}_{m{\bf k}\sigma})\langle \nabla_{\bf k}u^{(\xi)}_{n{\bf k}\sigma}|u^{(\xi)}_{m{\bf k}\sigma} \rangle$ and $\langle u^{(\xi)}_{n{\bf k}\sigma}|\frac{\partial {\bf H}^{(\xi)}_{{\bf k} \sigma}
}{\partial\xi}|u^{(\xi)}_{m{\bf k}\sigma} \rangle = (\epsilon^{(\xi)}_{n{\bf k}\sigma}-\epsilon^{(\xi)}_{m{\bf k}\sigma})\langle \partial_{\xi} u^{(\xi)}_{n{\bf k}\sigma}|u^{(\xi)}_{m{\bf k}\sigma} \rangle$ for $n\ne m$. These expressions can be obtained by taking the ${\bf k}$($\xi$)-gradient of ${\bf H}^{(\xi)}_{{\bf k} \sigma}|u^{(\xi)}_{n{\bf k}\sigma} \rangle=\epsilon^{(\xi)}_{n{\bf k}\sigma}|u^{(\xi)}_{n{\bf k}\sigma} \rangle$ and taking the inner product with $\langle u^{(\xi)}_{m{\bf k}\sigma}|$. Here, $|\psi^{(\xi)}_{n{\bf k}\sigma}\rangle=e^{i{\bf k}.{\bf r}}|u^{(\xi)}_{n{\bf k}\sigma}\rangle$ are the Bloch states and the summation is performed over the occupied bands.
The right hand side of  \eref{eq:pol12} is the Berry curvature which is gauge invariant and thus observable. \cite{Xiao10} The Berry curvature is nonzero in a wide range of materials, in particular, in crystals with broken time-reversal or inversion symmetry. In the periodic gauge \cite{periodicgauge}, the integral on $\xi$ can be done analytically and it leads to a two-point formula for the electric polarization that involves only the initial and final states of the system \cite{King-Smith93}:

\begin{eqnarray}
\Delta {\bf P}^{(\rm el)} &=&{\bf P}^{(\rm el)}(\xi_f)-{\bf P}^{(\rm el)}(\xi_i), \nonumber \\
{\bf P}^{(\rm el)}(\xi)&=&\frac{ie}{N}\sum_{{\bf k}\sigma}\sum_{n \in occ} 
\langle u^{(\xi)}_{n{\bf k}\sigma}|
\nabla_{\bf k}|u^{(\xi)}_{n{\bf k}\sigma} \rangle. 
\label{eq:pol3}
\end{eqnarray}

\paragraph{Ionic Hubbard model}
In 1D-IHM, the energy bands are $\epsilon_{\pm}(k)=\pm \epsilon(k)=\pm[\Delta^2+4t^2\cos^2(ka/2)+4\xi^2t^2\sin^2(ka/2)]^{1/2}$ and the bare current vertex and the bare distortion vertex in the band representation are given by  
$(\partial {\bf H}_0^{(\xi)}(k)/\partial k)={\bf U}^{-1}(k)[-at\sin(ka/2){\bm \tau}_x+a\xi t\cos(ka/2){\bm \tau}_y]{\bf U}(k)$
and 
$(\partial {\bf H}_0^{(\xi)}(k)/\partial \xi)={\bf U}^{-1}(k)[2 t\sin(ka/2){\bm \tau}_y]{\bf U}(k)$, respectively. ${\bm \tau}$ are Pauli matrices. ${\bf U}(k)$ and ${\bf U}^{-1}(k)$ diagonalize the IHM Hamiltonian as ${\bf H}_0(k)={\bf U}(k){\bm \epsilon}(k){\bf U}^{-1}(k)$ and are given by
\begin{widetext}
\begin{eqnarray}
{\bf H}_0(k) &=& [-2t\cos(ka/2){\bm \tau}_x+2\xi t\sin(ka/2){\bm \tau}_y-\Delta{\bm \tau}_z],\\
{\bf U}(k)&=&\frac{1}{\sqrt{2\epsilon(k)(\epsilon(k)-\Delta)}}[2i\xi t\sin(ka/2) {\bm 1}
+(\epsilon(k)-\Delta){\bm \tau}_x+2t\cos(ka/2){\bm \tau}_z],\\
{\bf U}^{-1}(k)&=&\frac{1}{\sqrt{2\epsilon(k)(\epsilon(k)-\Delta)}}[-2i\xi t\sin(ka/2) {\bm 1}
+(\epsilon(k)-\Delta){\bm \tau}_x+2t\cos(ka/2){\bm \tau}_z].
\end{eqnarray}
\end{widetext}
At small $\xi$ the denominator of \eref{eq:pol11} is given by $(\epsilon^{(\xi)}_{-k\sigma}-\epsilon^{(\xi)}_{+k\sigma})^2\simeq 4(\Delta^2+4t^2\cos^2(ka/2))$ while the imaginary part of the numerator is  
$(\partial {\bf H}_0^{(\xi)}/\partial k)_{12}(\partial {\bf H}_0^{(\xi)}/\partial \xi)_{21}-(\partial {\bf H}_0^{(\xi)}/\partial \xi)_{12}(\partial {\bf H}_0^{(\xi)}/\partial k)_{21}\simeq 2t^2\sin^2(ka/2)\cdot (\Delta/\sqrt{\Delta^2+4t^2\cos^2(ka/2)})$. Therefore, the polarization at small $\xi$ is given by

\begin{equation}
P^{(\rm el)}_{\Delta,U}(\xi) \simeq -\frac{2ea}{N}\sum_k\frac{ (\Delta/t)\sin^2(ka/2)}{[(\Delta/t)^2+4\cos^2(ka/2)]^{3/2}}\xi.
\label{eq:sd}
\end{equation} 

\section{Vertex corrections}\label{App:AppendixD}
The formula we derived in Sec.~ II involves vertex corrections at zero momentum transfer. The vertex corrections can be evaluated within DMFT in two different ways. A direct way uses the impurity model to compute the impurity two particle-hole irreducible vertex,  $\bm{\Gamma}_{irr}$, which is then used in the Bethe-Salpeter equation graphically shown in \fref{feyndiag2}. \cite{Toschi07, Rohringer12} 

Alternatively, the vertex function of a current $A =\sum_{{\bf k}\sigma\sigma'}\lambda^A_{{\bf k}\sigma\sigma'}c^{\dagger}_{{\bf k}\sigma}c_{{\bf k}\sigma'}$ is calculated by adding $\zeta A$ to the Hamiltonian where $\zeta$ is a small parameter, and measure the self-energy variation relative to the field, $\partial {\bm \Sigma}/\partial \zeta$. Note that due to diagonal self-energy in DMFT approximation, the vertex functions are diagonal as well. 

We are interested in the current and distortion vertices which are described by the following operators:
\begin{eqnarray}
\bm{\lambda}^{(\xi)}_{j}({\bf k})&=&at\sin(ka/2){\bm \tau}_x+a\xi t\cos(ka/2){\bm \tau}_y,\label{eq:jvertex}\\
\bm{\lambda}^{(\xi)}_{H'}({\bf k})&=&2 t\sin(ka/2){\bm \tau}_y.\label{eq:Hvertex}
\end{eqnarray}

\begin{figure}
\begin{center}
\includegraphics[width=0.55\columnwidth]{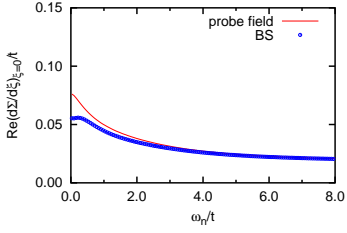} 
\caption{(Color online) The real part of the distortion vertex function, $(\partial{\bm \Sigma}/\partial\xi)_{\xi=0}$, as a function of Matsubara frequency at $U=5.0t$ obtained from the two methods described in the text. }
\label{dsigma}
\end{center}
\end{figure}

We have evaluated the vertex functions corresponding to \eref{eq:Hvertex} using the two methods, and we display them in \fref{dsigma} for $U/t=5.0$. The small discrepancy between two data sets is due to 
lesser number of the bath levels and frequency cutoff in evaluating $\bm{\Gamma}_{irr}$. This introduces numerical errors once we evaluate the fully reducible vertex function for the lattice. The current vertex function is identically zero. 

It is not clear from the Bethe-Salpeter equation why the current vertex function is zero. Next we discuss this in more details. In the DMFT, the lattice irreducible vertex function, ${\bm \Gamma}_{irr}$,  is approximated by the impurity irreducible vertex function which is purely local. Thus, the different momentum vectors appearing on both sides of the irreducible vertex function can be summed over independently, ignoring momentum conservation at the vertex $\bm{\Gamma}_{irr}$. \cite{Pruschke93} Furthermore, since we are defining the two impurity model for two sites within the unit cell, ${\bm \Gamma}_{irr, l_1l_2,l_3l_4}$ is diagonal in orbital indices within unit cell, i.e., it is nonzero only if $l_1=l_2=l_3=l_4$.  Note that ${\bm \Gamma}_{irr}^{\xi}$ is basically a function of $\xi$ and should be evaluated for each value of $\xi$.  

Having ${\bm \Gamma}_{irr}^{\xi}$ one can explicitly calculate the distortion vertex corrections. We need to close two legs of the irreducible vertex function with the following function to obtain the vertex corrections at the lowest order (however, the following argument is valid for all orders)
\begin{equation}
F_{ll',\sigma}^{j_x(H')}(\xi,i\omega_m)=\frac{1}{N}\sum_k\sum_{l_1l_2}{\bf G}_{ll_1,k\sigma}{\bm \lambda}_{l_1l_2}^{j_x(H')}(k){\bf G}_{l_2l',k\sigma}.
\end{equation}

Although in general all elements of ${\bf F}$ are necessary, due to locality of ${\bm \Gamma}_{irr}$ in orbital indices we only need to calculate diagonal elements. Furthermore, since only nondiagonal elements of the bare vertices are nonzero, and  $\lambda_{12}=\lambda_{21}^*$, we can write
\begin{equation}
F_{11,\sigma}^{j_x(H')}=(1/N)\sum_k G_{11,\sigma}\Re (\lambda_{12}^{j_x(H')} G_{21,\sigma}),
\end{equation}
where we have used the identity $G_{12}=G_{21}^*$ 
. The interacting Green's function of the ionic-Hubbard model is given by 
\begin{multline}
{\bf G}(k,i\omega_m) = [(i\omega_m+\mu){\bm 1} 
-\Delta{\bm \tau}_z-{\bm \Pi}(i\omega_m)\\-2t\cos(ka/2){\bm \tau}_x-2\xi t\sin(ka/2){\bm \tau}_y]/E(k,i\omega_m)
\end{multline}
where in DMFT approximation ${\bm \Pi}(i\omega_m)$ is a diagonal matrix with ${\bm \Pi}_{11}=\Sigma_B$ and ${\bm \Pi}_{22}=\Sigma_A$  and 
\begin{multline}
E(k,i\omega_m) = -4t^2\cos^2(ka/2)-4\xi^2 t^2 \sin^2(ka/2)+\\
[i(\omega_m+\Im\Sigma_A)+\mu+\Delta-\Re\Sigma_A][i(\omega_m+\Im\Sigma_B)+\mu-\Delta-\Re\Sigma_B].
\end{multline}
Due to the particle-hole symmetry of the model, we have $\Im\Sigma_A=\Im\Sigma_B$ 
and $\mu+\Delta-\Re\Sigma_A=-(\mu-\Delta-\Re\Sigma_B)$.

Substituting ${\bf G}(k,i\omega_m)$ in the definition of $F_{11}$, it can be easily shown that the momentum dependency of the $F_{11}$ has the following form for the current and the distortion vertices: 
\begin{eqnarray}
G_{11,\sigma}\Re (\lambda_{12}^{j_x} G_{21,\sigma})&\propto&\frac{\sin(ka/2)\cos(ka/2)}{E(k,i\omega_m) } \label{eq:cvc}\\ 
G_{11,\sigma}\Re (\lambda_{12}^{H'} G_{21,\sigma})&\propto& \frac{\sin^2(ka/2)}{E(k,i\omega_m)}.\label{eq:dvc}
\end{eqnarray}

Since $E(k,i\omega_m)$ is an even function of $k$, in the momentum-summation the current vertex corrections, \eref{eq:cvc}, vanish ($F_{22}$ has same momentum dependency). 

From a direct calculation of the electric polarization with and without distortion vertex corrections we found that the vertex corrections contribution in the electric polarization is negligible in particular at the small to intermediate interaction strengths. For largest interaction strength and the largest distortion value considered in \fref{fig2} the vertex corrections contribution in the electric polarization is less than $\sim 8\%$.
 Vertex corrections introduce  changes in the response functions due to the multiple scattering
of  real or virtual  particle-hole excitations.  In an insulating state at low temperatures,   this 
correction is inversely proportional to the gap, and it is small far from the metal to insulator
transition.

\bibliographystyle{prsty}

\begin{thebibliography}{10}
\bibitem{BO}
In a centrosymmetric system of interacting electrons, the variation of the polarization as a function of $U$ depends on whether spontaneous breaking of inversion symmetry happens or not. A DMFT calculation cannot capture a spontaneous inversion symmetry broken phase and Cluster-DMFT breaks translational symmetry by construction, so for the centrosymmetric system we only consider the symmetric Mott phase.
\end{thebibliography}

\end{document}